\documentclass[twocolumn,twoside,slac]{revtex4}
\usepackage{graphicx,rotating}
\usepackage{fancyhdr}
\usepackage{epsf}
\begin{document}

\title{ POOL File Catalog, Collection and Metadata Components }

%
%
\author{ C. Cioffi }
\affiliation{University of Oxford,  Oxford, OX13NP, UK}
\author{ S. Eckmann }
\affiliation{Argonne National Laboratory, Argonne, IL 60439, USA}
\author{ M. Girone }
\affiliation{CERN, 1211 Geneve 23, Switzerland}
\author{ J. Hrivnac }
\affiliation{ LAL, Orsay, France}
\author{ D. Malon }
\affiliation{Argonne National Laboratory, Argonne, IL 60439, USA}
\author{ H. Schmuecker }
\affiliation{CERN, 1211 Geneve 23, Switzerland}
\author{ A. Vaniachine }
\affiliation{Argonne National Laboratory, Argonne, IL 60439, USA}
\author{ J. Wojcieszuk }
\affiliation{CERN, 1211 Geneve 23, Switzerland}
\author{Z. Xie\footnote{current address: CERN, 1211 Geneva 23, Switzerland} }
\affiliation{Princeton University, Princeton, NJ 08544, USA}

\begin{abstract}
 The POOL project is the common persistency framework for the LHC experiments to store petabytes of experiment data and metadata in a distributed and grid enabled way. POOL is a hybrid event store consisting of a data streaming layer and a relational layer.

 This paper describes the design of file catalog, collection and metadata components which are not part of the data streaming layer of POOL and outlines how POOL aims to provide transparent and efficient data access for a wide range of environments and use cases - ranging from a large production site down to single disconnected laptops.

 The file catalog is the central POOL component translating logical data references to physical data files in a grid environment. POOL collections with their associated metadata provide an abstract way of accessing experiment data via their logical grouping into sets of related data objects. 

\end{abstract}
\maketitle
\thispagestyle{fancy}
\section{Introduction}

The POOL project\cite{pool} is the common persistency framework for the LHC experiments to store petabytes of experiment data and metadata in a distributed and grid enabled way. The POOL is a hybrid event store combining C++ object streaming technology such as Root I/O \cite{root} for the bulk data with a transactionally safe Relational Database store such as MySQL for file catalog, collection and metadata.

 This paper describes the design of POOL components which are not part of the data streaming layer\cite{stream} and outlines how POOL aims to provide transparent and efficient data access for a wide range of environments and use cases - ranging from a large production site down to single disconnected laptops.

 The POOL file catalog is the central POOL component translating logical data references to physical data files in a grid environment. POOL collections with their associated metadata add another more abstract way of accessing experiment data via their logical grouping into sets of related data objects. Ad hoc queries on the collections provide physicist with an efficient way to extract useful data.

\section{File catalog component}
\subsection{Purpose}
 In a file based persistency mechanism such as POOL, the storage components need to operate on the contents of the file. One should be able to navigate from one object to the other even if they are not stored in the same file. Using an external file catalog to keep track of the physical location of the file is more flexible than hard-coding the location information in the file itself. It allows the file to be moved or replicated. 

 In the grid environment, high level applications view only logical files. A logical file does not exist physically. A physical file and all its copies can be viewed as the same logical file. In another word, physical files are representations of the logical file. To operate on the contents of a logical file, a service is needed to map the logical file to one of its physical representations and this physical file can then be opened. 

The physical location of a file can be represented by its Physical File Name(PFN) while the logical file can be represented by a unique logical identifier. In POOL, this identifier is called the file ID and is generated via the GUID (Global Unique Identifier) mechanism.      

The file catalog component in POOL is responsible for maintaining a list of physical locations of accessible files together with their unique and immutable IDs and translating the logical file reference into its physical representation. It is a basic component for object navigation and grid integration. 
  
\subsection{Design and implementation}

\begin{figure*}[tb]
\vspace{8em}
\hspace{-8em}
\begin{rotate}{-90}
  \epsfxsize=100mm
\epsffile[200 300 895 842]{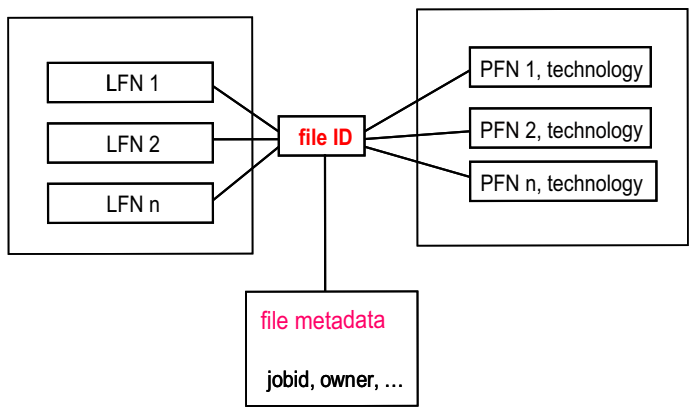}
\end{rotate}
\vspace{10em}
\caption{Logical view of the catalog }
\label{fig:FCSchema}
\end{figure*}

The basic content of the catalog is the one-to-many mapping of the file ID and the PFNs. This mapping is sufficient for object lookup and navigation. Optionally, the file catalog can store logical file names(LFNs). In contrast to the file ID, which is not easy for the user to read and to remember, the LFNs may contain human readable and memorisable strings. LFN can be seen as the alias of the file ID. One file ID may have many LFNs. In addition to PFN and LFN, a set of user defined file metadata can also be associated with a file ID. Metadata-fileID mapping is one-to-one which means one file can be associated with one set of well defined metadata. The purpose of the file metadata support is to help selecting a fragment of the catalog when necessary. Figure \ref{fig:FCSchema} shows the logical view of the content of the catalog: the PFN-fileID mapping, which is essential for the object lookup and optionally, the LFN-fileID and the metadata-fileID mapping.   

File registration and lookup are two fundamental functionalities of the component. To register a file is to insert a file ID and PFN pair in the catalog while to lookup a file is to resolve a file ID into a PFN or vice versa. Similar functions are provided also for the file ID and LFN mapping. Besides, one can extract a subset of files from one concrete catalog and cross populate them into another catalog. Specific metadata can be associated to the file and queries on the metadata can be used to select the catalog fragments. In the cross catalog operations the source and the destination catalogs need not to have the same backend implementation.

 The component provides both C++ API and command-line tools. The C++ API is used by the storage components in POOL and the experiment framework to register and lookup a file inside the application process while command-line tools can be used outside the application process for catalog management operations. 

Three concrete catalog: XML, MySQL and EDG-RLS based implementations are provided under single abstract interface. Concrete catalogs are loaded dynamically at run time. To take advantage of the relational database technology, transactions are supported by the component. In the transaction the user can commit or rollback the changes to the catalog. 

The three implementations are described in more detail as follows:   

\begin{itemize}
\item XML catalog
 
The XML catalog is useful when the user wants to run the application disconnected from the network and the number of entries in the catalog is not too large. Running applications connected to a local XML catalog do not rely on any central service. The XML catalog can also be used in data migration. For example, when one migrates the data in a MySQL backend into a, say, Oracle backend, the XML catalog can be used as the intermediate format.  
  
\item MySQL catalog

The MySQL catalog is connected to the MySQL database server. The database based catalog can handle concurrent accesses and larger data volume than the XML catalog. The MySQL catalog can be used in larger scale applications such as in a production farm.  

\item EDG catalog

A Grid based catalog can be used by the entire Virtual Organisation (VO). The EDG project \cite{edg} will provide the Replica Management Service, which controls files that belong to a VO. In particular, the Replica Location Service(RLS) component\cite{rls} maintains information about the physical location of files, while the Replica Metadata Catalog(RMC) component\cite{rmc} provides the information on the logical file names and metadata. The file catalog component in POOL provides an interface to the EDG-RLS and EDG-RMC. In another word, POOL applications connected to the EDG catalog are on the Grid.   

\end{itemize}

The file catalog component provides a Graphic User Interface for browsing the content of the catalog. A first prototype of the browser has been developed in Python scripting language. 
%

\subsection{Example use cases}
\begin {itemize}
\item Read/Write files in POOL applications

 When the POOL persistency manager writes in a file, it requests the ID of the file of given PFN from the file catalog component. The PFN may or may not already be registered in the catalog. If the file is already registered, the file catalog component returns the file ID to the persistency manager, otherwise the file catalog component generates a new file ID, registers it in the catalog and returns it to the persistency manager.

 When the persistency manager opens a file for reading, it requests the PFN of the file of a given file ID from the file catalog component. The file catalog component looks up the PFN in the catalog and returns it to the persistency manager. If more than one PFNs are found, in the case of a Grid based catalog, the optimal PFN will be returned, in an off-the-Grid case, the PFN of the master copy of the file will be returned.         
  
\item Manage a production in a local farm  

The user extracts a catalog fragment needed by the job for reading from a central catalog(EDG or MySQL based) into a local XML catalog. User runs the job disconnected from the network. During the running of the job, the local input XML catalog is used when retrieving objects; the output files are registered into another local XML catalog. After n job runs, n output XML catalogs are produced. User cleans the entries in the catalogs produced by unsuccessful jobs and publishes the local catalog fragments to the central catalog making the produced data files available to a group of users or the entire production site. 

In this setup, when jobs are running they do not depend on any central services.  
\end {itemize}

\subsection{ Preliminary performance tests }

Performance tests have been set up to compare the behavior of different catalog implementations. 

For the XML catalog, tests were run on a single Pentium III-1.2GHz computer with about 200MB free memory. The maximum number of entries have been tested in a single XML catalog is 50,000. The major time spent for the XML catalog is on the XML DOM parser initialization. It takes about 10ms to initialize a new catalog and about 6s to start a catalog with 20K entries. The PFN registration operation takes about 0.3 ms/entry on average.  

For the MySQL catalog, tests were run on 10 client nodes and a MySQL server node on the LAN. The maximum number of entries have been tested in a single MySQL catalog is 1 million. In the case of 300 concurrent client processes from the 10 client nodes querying the MySQL server and committing every 100 entries, the PFN registration operation takes about 1.5 ms/entry on average.

Similar tests have been performed against the EDG-RLS server which is web service on top of relational database. The PFN registration time is in the range from about 30ms/entry to 6ms/entry for multithreaded applications depending on the combination and the setup of the web service and the backend database.

In all tests, a single PFN entry is about 200 characters long and the file ID size is 36 characters.

This is the first performance test of the POOL file catalog, no systematic performance test and tuning have been attempted so far. 

\section{Collection and Metadata component}

\subsection{Overview}

The purpose of the POOL collection component is to 
provide the tools needed to manage potentially large ensembles 
of objects stored by means of POOL's persistence services. 
The collection component provides the infrastructure to support 
definition, creation, population, and use of such ensembles, 
including query, selection, and iteration services, and higher-
level utilities. Collections provide an entry point for 
physicists, a locus for recording the navigational information 
necessary to allow later iteration through a list of data 
objects that may be physically dispersed across a very large persistent 
store. A collection of events that are of interest to a particular 
physics working group because they share certain physics 
characteristics is a motivating example.  

To support selection of specific subsets of interest from  
very large collections, it is useful to provide a means to 
query attributes of those objects without retrieving the 
objects themselves. The metadata component in POOL provides 
the machinery to define and manage object- and collection-level 
metadata. 

Named collections populated by explicit calls to insertion 
operators are called ``explicit collections'' in POOL.  
POOL also supports ``implicit collections''--collections of objects 
grouped by physical containment. The set of all objects of type T 
in a given container or file is an example of an implicit collection.
An object persistified by writing its state into a file is implicitly 
an element of the collection of all such objects 
in the file, but the same object may also be registered in one 
or more explicit collections.
  
Explicit and implicit collections are read via the same abstract 
interface, so client code does not in any way depend upon whether 
an input collection is explicit or implicit.  

The first implementation of explicit collections is in MySQL.
The intent is to allow the possibility of server-side selection 
of objects of interest from queryable collections by means of 
database-provided SQL services, so that the list of qualifying 
objects, and, by extension, the list of data files needed, may be 
determined at a stage prior to job scheduling and grid resource 
brokering, without the need to first retrieve the data files 
and navigate into application-specific data.

POOL also supports hierarchical collections, tree-like structures 
that accept other (sub)collections as their elements.
The design allows complex selection based upon both collection- 
and object-level metadata.

\subsection{Example use cases}

\begin{itemize}
\item Create/Write explicit collections

 User first defines the attribute lists of collections will be produced by the process.
 Then she processes the event and gathers attribute data. For each event, the attribute data and a reference, which contains enough information to retrieve the persistent object, is added to the corresponding collection. At the end of process, several collections with different names are created and persisted.    
 
\item Read explicit collections

During analysis, the user specifies a number of collections and provides a query to select only the objects of interest. Then she iterates through the returned persistent object references, retrieve the objects and process them in the analysis. 

\end{itemize}

\section{Summary}

File catalog, Collection and Metadata components are part of the so-called relational layer of the hybrid event store POOL. Such a relational layer takes advantage of the underlying relational database technology, e.g. the transaction, concurrency and queries, to achieve efficient management of files and object metadata. 

The file catalog is one of the basic components which provide transparent object navigation in the event store. The design of the component can accommodate wide ranges of use cases. It is grid-aware but also preserves grid-decoupled modes. 

The collection and metadata components provide physicists with higher level access to and efficient selection on the experiment data in the event store.    
 


\end{document}